\documentclass[english,reprint,aps,prx,superscriptaddress,longbibliography]{revtex4-2}
\usepackage[T1]{fontenc}
\usepackage[latin9]{inputenc}
\usepackage{amsmath}
\usepackage{amssymb}
\usepackage{graphicx}

\makeatletter

\pdfpageheight\paperheight
\pdfpagewidth\paperwidth

\usepackage{color}
\usepackage{hyperref}
\hypersetup{
  colorlinks=true,
  citecolor=blue,
  linkcolor=blue,
  urlcolor=blue,
}

\makeatother

\usepackage{babel}
\begin{document}
\title{Light-induced phase transitions in vanadium dioxide: a tensor network study}

\author{Lin Zhang}
\email{lin.zhang@icfo.eu}
\affiliation{ICFO-Institut de Ciencies Fotoniques, The Barcelona Institute of Science and Technology, Castelldefels (Barcelona) 08860, Spain}

\author{Utso Bhattacharya}
\email{ubhattachary@phys.ethz.ch}
\affiliation{ICFO-Institut de Ciencies Fotoniques, The Barcelona Institute of Science and Technology, Castelldefels (Barcelona) 08860, Spain}
\affiliation{Institute for Theoretical Physics, ETH Zurich, 8093 Zurich, Switzerland}

\author{Maria Recasens}
\affiliation{ICFO-Institut de Ciencies Fotoniques, The Barcelona Institute of Science and Technology, Castelldefels (Barcelona) 08860, Spain}

\author{Tobias Grass}
\affiliation{DIPC - Donostia International Physics Center, Paseo Manuel de Lardiz{\'a}bal 4, 20018 San Sebasti{\'a}n, Spain}
\affiliation{IKERBASQUE, Basque Foundation for Science, Plaza Euskadi 5, 48009 Bilbao, Spain}

\author{Ravindra W. Chhajlany}
\affiliation{Institute of Spintronics and Quantum Information, Faculty of Physics, Adam Mickiewicz University, 61614 Poznan, Poland}

\author{Maciej Lewenstein}
\affiliation{ICFO-Institut de Ciencies Fotoniques, The Barcelona Institute of Science and Technology, Castelldefels (Barcelona) 08860, Spain}
\affiliation{ICREA, Pg. Lluis Companys 23, 08010 Barcelona, Spain}

\author{Allan S. Johnson}
\email{allan.johnson@imdea.org}
\affiliation{IMDEA Nanoscience, Calle Faraday 9, 28049, Madrid, Spain}

\begin{abstract}
Nonequilibrium phase transitions driven by light pulses represent a rapidly developing field in condensed matter physics
as they offer an efficient way to tune and control material properties on ultrafast timescales. As one of the
archetypal strongly correlated materials, vanadium dioxide ($\mathrm{VO}_{2}$) undergoes a structural phase transition (SPT)
from a monoclinic (M1) to rutile (R) structure and an insulator-to-metal transition (IMT) either when heated above 340~K or when excited by an ultrafast laser pulse. Here, we present a tensor network study of the light-induced phase transitions in $\mathrm{VO}_{2}$ based on a quasi-one-dimensional model with all the important ingredients---multi-orbital character, electron-lattice coupling, and electron-electron correlations---being included. We show that this model qualitatively captures the equilibrium properties of $\mathrm{VO}_{2}$ by calculating the ground state phase diagram and finite-temperature phase transitions. A hybrid quantum-classical tensor-network method is used to simulate the dynamics following photoexcitation. We find that the structure can transform faster than the harmonic phonon modes of M1 phase, suggesting lattice nonlinearity is key in the SPT. We also find separate timescales for the evolution of dimerization and tilt distortions in the lattice dynamics, as well as the loss and subsequent partial restoration behavior of the displacements, which can provide an explanation for the complex dynamics observed in recent experiments [C. Brahms {\it et al.}, arXiv:2402.01266]. Moreover, decoupled SPT and IMT dynamics are observed in the numerical simulations: while the initial M1 structure transforms to the R one in tens of femtoseconds, the IMT occurs quasi-instantaneously, consistent with recent experimental findings. Our theoretical studies provide insight into the light-induced phase transitions of $\mathrm{VO}_{2}$, revealing unexpected non-monotonic transformation pathways and paving the way for future studies of non-thermal phase transformations.

\end{abstract}
\maketitle

\section{Introduction}

One of the archetypal strongly correlated materials, vanadium dioxide ($\mathrm{VO}_{2}$) is a transition-metal compound which undergoes a first-order transition from the insulating phase to the metallic phase at $T_{\mathrm{c}}\approx 340\,\mathrm{K}$ and ambient pressure~\citep{Mott1949,Morin1959,Liu2018,Shao2018}. Coinciding with this insulator-to-metal transition (IMT), a structural phase transition (SPT) also occurs from the low-temperature distorted monoclinic (M1) phase to the high-temperature undistorted rutile (R) stucture~\citep{Andersson1954,Andersson1956,Goodenough1960}. Due to the strong correlations between the internal charge, orbital and lattice degrees of freedom, the underlying mechanism of these transitions in $\mathrm{VO}_{2}$ is still under debate~\citep{Biermann2005,Eyert2011,Brito2016,Najera2017,Najera2018,Kim2013}. In particular, it remains unclear whether the transition is best described as a Peierls-like transition driven by the structure change of lattice~\citep{Goodenough1971} or as Mott-like transition driven by the electron-electron correlations~\citep{Zylbersztejn1975}.

On the other hand, nonequilibrium phase transitions in materials induced by ultrafast light pulses are  attracting considerable attention and represent a rapidly developing field in condensed matter physics~\citep{Nasu2004,Fausti2011,Giannetti2016,Torre2021,Koshihara2022,Rajpurohit2022}, as they offer an efficient way to tune and control material properties on ultrafast timescales. In $\mathrm{VO}_{2}$, intense laser pulses can suddenly change the potential energy surface of lattice through electronic excitation and drive the ultrafast SPT and IMT~\citep{Cavalleri2004,Wall2012}. As in principle the lattice and electronic degrees of freedom can respond on different timescales, the light-induced phase transition has become one of the key tools to address the nature of IMT~\citep{Baum2007,Kuebler2007,Liu2012,Cocker2012,Tao2012,Morrison2014,Wegkamp2015,OCallahan2015,Li2017,Wall2018,Otto2018,Lee2018,Fu2020,Vidas2020,Sood2021,Johnson2022}. Early experiments highlighted the role of lattice distortions in the light-induced IMT~\citep{Cavalleri2001,Cavalleri2004}, but more recent studies suggest that the IMT is  faster than the SPT and emphasized the importance of electron-electron correlations~\citep{Pashkin2011,Wegkamp2014,Jager2017,Bionta2018}. In a recent experiment~\citep{Brahms2024}, the complete structural and electronic nature of light-induced phase transitions in $\mathrm{VO}_{2}$ have been resolved  at their fundamental time scales using ultra-broadband few-femtosecond spectroscopy. In addition to a quasi-instantaneous IMT, a much more complex pathway in the light-induced phase transitions was observed. 

However, in contrast to these experimental advances, there have been very limited theoretical studies on the nonequilibrium phase transition in $\mathrm{VO}_{2}$. This is naturally due to the complexity in treating even the normal thermal transition in $\mathrm{VO}_{2}$, and so most studies have used simplified static models~\citep{Yuan2013,Wegkamp2014,Jager2017} or structural only models~\citep{Wall2018} to interpret the transient signatures. Only recently has time-dependent density functional theory (TD-DFT) been applied to the problem~\citep{Xu2022,Liu2022}, but the use of  DFT to describe $\mathrm{VO}_{2}$ has often been controversial due to the neglect of electron-electron interactions. The complexity of uncovering the important couplings from DFT has also motivated the use of simplified models in the past~\citep{Grandi2020}. Furthermore, these works predict transformation times that are strongly dependent on the excitation fraction and initial temperature, an effect not seen in recent ultrafast X-ray diffraction studies~\citep{Wall2018,Johnson2023,DeLaPenaMunoz2023}. 

To overcome these limitations and provide a more transparent model, here we present a tensor network study of the light-induced phase transitions using a simplified quasi-one-dimensional model for $\mathrm{VO}_{2}$, taking into account for the first time all the important physical ingredients: the multiorbital character, electron-lattice coupling, and electron-electron correlations. We show that this model qualitatively captures the equilibrium properties of $\mathrm{VO}_{2}$ by calculating the ground state phase diagram and finite-temperature phase transitions. When the light pulse is applied to the system, a hybrid quantum-classical tensor-network method is used to simulate the dynamics. We find that the structure can transform faster than the corresponding harmonic phonon modes of M1 phase, suggesting lattice nonlinearity is key in the SPT.  We also find separate timescales for the evolution of dimerization and tilt distortions in the lattice dynamics, and that the displacements exhibit a loss and subsequent partial restoration behavior, which can provide an explanation for the complex dynamics observed in Ref.~\citep{Brahms2024}. Moreover, decoupled SPT and IMT dynamics are observed, where the initial M1 structure transforms to the R one in tens of femtoseconds, while the IMT occurs quasi-instantaneously. Our results support the recent experimental findings and provide key insights into the light-induced phase transitions in $\mathrm{VO}_{2}$.

The remaining part of this article is organized as follows. In Sec.~\ref{sec:Model}, we introduce the quasi-one-dimensional model. Then in Sec.~\ref{sec:Equilibrium properties} we show that this model qualitatively captures the essential equilibrium physics of $\mathrm{VO}_{2}$. The light-induced phase transitions are studied in Sec.~\ref{sec:Light-induced phase transitions}. Finally, we present the conclusion in Sec.~\ref{sec:Conclusion}. More details are provided in the appendices.

\begin{figure}
\includegraphics[scale=0.7]{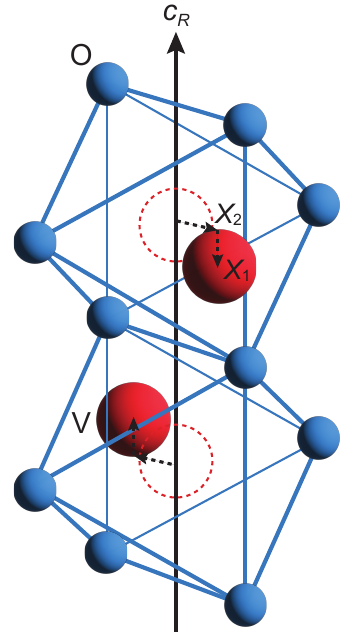}

\caption{Crystal structure of $\mathrm{VO}_{2}$. Here the red (blue) spheres represent
vanadium (oxygen) atoms. In the rutile R phase, the vanadium atoms are located at the positions of dashed circles. 
The finite $X_{1}$ and $X_{2}$ lattice distortions characterize the monoclinic M1 phase, 
where the $X_{1}$ component captures the dimerization along the $c_{R}$ axis and 
the $X_{2}$ component acts as a tilting perpendicular to the $c_{R}$ axis. \label{fig:crystal structure}}
\end{figure}

\section{Model\label{sec:Model}}

Our model for $\mathrm{VO}_{2}$ is inspired by the earlier static model of Ref.~\citep{Grandi2020}. In $\mathrm{VO}_{2}$, the vanadium 3d orbitals hybridize and split under the action of the crystal field. The relevant orbitals to the IMT and SPT are the $a_{1g} $ singlet and $e_{g}^{\pi}$ doublet (equivalently and often referred to as such in the literature, $d_{||}$ singlet and $\pi^*$ doublet). Like Ref.~\citep{Grandi2020}, we consider only one $e_{g}^{\pi}$ orbital to
simplify the theoretical model without losing the important
physics of $\mathrm{VO}_{2}$. Since the Peierls instability mainly
occurs along the $c_{R}$ axis connecting adjacent vanadium ions, we model the vanadium dioxide as a quasi-one-dimensional
system, for which the lattice displacement $\mathbf{X}\equiv(X_{1},X_{2})$
is introduced to capture the dimerizing displacement along the $c_{R}$
axis and the band-splitting tilting displacement perpendicular to the $c_{R}$ axis,
respectively; see Fig.~\ref{fig:crystal structure}. The total Hamiltonian for this simplified model of $\mathrm{VO}_{2}$ with the coupling to lattice degrees
of freedom is given by
\begin{equation}
H=H_{\mathrm{e}}+H_{\text{e-}\mathbf{X}}+\Phi(\mathbf{X}).\label{eq:Hamiltonian}
\end{equation} As we will shortly show, we treat the electronic component fully quantum mechanically, while treating the nuclei classically, leading to a ``semi-quantum'' approach.

The purely electronic component reads
\begin{equation}
\begin{aligned}H_{\mathrm{e}}= & -\sum_{i}\sum_{a=1,2}\sum_{\sigma=\uparrow,\downarrow}t_{a}c_{a,\sigma,i}^{\dagger}c_{a,\sigma,i+1}\\
 & -t_{12}\sum_{i}\sum_{\sigma=\uparrow,\downarrow}c_{1,\sigma,i}^{\dagger}c_{2,\sigma,i}+\mathrm{H.c}\\
 & +\sum_{i}\sum_{a=1,2}\varepsilon_{a}n_{a,i}+\frac{U}{2}\sum_{i}n_{i}(n_{i}-1),
\end{aligned}
\end{equation}
where $a=1,2$ denotes the $a_{1g}$ and $e_{g}^{\pi}$ orbital, respectively,
and $c_{a,\sigma,i}$ is the annihilation operator for electron at site $i$ with 
orbital $a$ and spin $\sigma$. The nearest-neighbor intra-orbital hopping is given by $t_{a}$,
while $t_{12}$ is the onsite inter-orbital hopping. Here, $\varepsilon_{a}$ and $U$ describe 
the onsite energy potential and Hubbard repulsive interaction, respectively. 
We have the particle number operator $n_{i}=\sum_{a=1,2}n_{a,i}$ and $n_{a,i}=\sum_{\sigma=\uparrow,\downarrow}n_{a,\sigma,i}$ with $n_{a,\sigma,i}\equiv c^{\dagger}_{a,\sigma,i}c_{a,\sigma,i}$.
The system is at quarter filling. 

The lattice distortion can be
modelled through the classical potential energy~\citep{Grandi2020}
\begin{equation}\label{eq:lattice potential energy}
\begin{aligned}\Phi(\mathbf{X})=L & \left[\frac{\alpha}{2}(X_{1}^{2}+X_{2}^{2})+\frac{\beta_{1}}{4}(2X_{1}X_{2})^{2}\right.\\
 & \left.+\frac{\beta_{2}}{4}(X_{1}^{2}-X_{2}^{2})^{2}+\frac{\gamma}{6}(X_{1}^{2}+X_{2}^{2})^{3}\right],
\end{aligned}
\end{equation}
which is obtained from the Landau functional for improper ferroelectrics
expanded up to the sixth order in the lattice displacements to accurately recover the first order nature of the transition. Here, $L$ is the number of lattice sites. The first term and the last term are fully rotationally symmetric in the $X_{1}$-$X_{2}$ plane. On
the other hand, the term proportional to $\beta_{1}$ favors a lattice
distortion only along one of these two directions, whereas the term
proportional to $\beta_{2}$ favors a distortion with $|X_{1}|=|X_{2}|$.
In the case of $\mathrm{VO}_{2}$, both the displacements $X_{1}$ and $X_{2}$
are nonzero (i.e., there are both dimerization and tilt in the displacements), hence we should have $\beta_{2}>\beta_{1}$.

Finally, for the electron-lattice coupling we have
\begin{equation}\label{eq:electron-lattice coupling}
H_{\text{e-}\mathbf{X}}=-gX_{1}\sum_{i}(-1)^{i}n_{1,i}-\frac{\delta}{2}X_{2}^{2}\sum_{i}(n_{1,i}-n_{2,i}).
\end{equation}
The first term describes the dimerization induced by
the displacement $X_{1}$ along the $c_{R}$ axis and is controlled by
the coupling constant $g$, while the second term with strength $\delta$
represents the crystal field splitting generated by the tilting displacement
$X_{2}$.  The coupling to $X_{1}$ is linear at leading order,
whereas the coupling to $X_{2}$ is quadratic since the opposite variations
of the hybridization between the $e_{g}^{\pi}$ orbital and the closer/further
oxygen ligands at linear order in $X_{2}$ cancel each other, but their
sum is nonzero at second order~\citep{Grandi2020}. Note that the
total Hamiltonian is invariant under the transformations $X_{1,2}\to-X_{1,2}$
and possesses a $Z_{2}\times Z_{2}$ symmetry.

We emphasize that, with the simplified quasi-one-dimensional model \eqref{eq:Hamiltonian}, 
our goal is to qualitatively reproduce the physics of $\mathrm{VO}_{2}$, especially the 
light-induced nonequilibrium phase transitions, without any ambition for quantitative agreement.
For this, like Ref.~\citep{Grandi2020} we assume that the bands for $a_{1g}$ and $e_{g}^{\pi}$ orbitals
have the same bandwidth and center of gravity (i.e., $\varepsilon_{1}=\varepsilon_{2}=0$) to reduce 
the number of Hamiltonian parameters. We  set the half-bandwidth to $1\,\mathrm{eV}$, i.e., $t_{1}=t_{2}=0.5\,\mathrm{eV}$, and the inter-orbital
hopping coefficient as $t_{12}=0.1\,\mathrm{eV}$, which is small compared
with the intra-orbital hopping. For the Hubbard interaction, we choose $U=0.6\,\mathrm{eV}$ to generate a zero-temperature
energy landscape that is similar to the one shown in Ref.~\citep{Grandi2020},
where the parameters for the lattice potential are set as $\alpha=0.155\,\mathrm{eV}$,
$\beta_{1}=1.75\times10^{-3}\,\mathrm{eV}$, $\beta_{2}=2\beta_{1}$, and $\gamma=6.722\times10^{-4}\,\mathrm{eV}$.
We use the same lattice potential parameters in this work. Finally,
we choose the electron-lattice coupling strength as $g=0.528\,\mathrm{eV}$ and
$\delta=0.2\,\mathrm{eV}$, such that the transition temperature from the M1 phase
to the R phase is close to the experimental value; see Sec.~\ref{subsec:phase transition at finite temperature}.

We note that our definition of lattice potential parameters and electron-lattice couplings in units of energy implies that the displacements $X_{1}$ and $X_{2}$ are expressed in a dimensionless way. The underlying length scale (on the order of $0.1\,\mathrm{\mathring{A}}$) is not relevant for distinguishing the R phase ($X_{1}=X_{2}=0$) and the M1 phase ($X_{1}\neq 0$ and $X_{2}\neq 0$).

\begin{figure}
\includegraphics{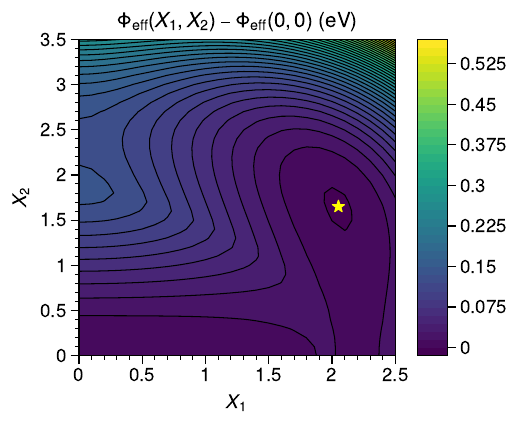}

\caption{The zero-temperature internal energy density $\Phi_{\mathrm{eff}}$ as a function
of the lattice distortions $X_{1}$ and $X_{2}$. Due to the $Z_{2}\times Z_{2}$
symmetry of the system, we only show the results for the region with $X_{1},X_{2}>0$,
where the internal energy has two minima, one located at $X_{1}=X_{2}=0$
corresponding to the undistorted R phase, and another located at $X_{1}\approx2.05$
and $X_{2}\approx1.65$ corresponding to the distorted M1 phase. Here
we have $\Phi_{\mathrm{eff}}(0,0)\approx-0.71804\,\mathrm{eV}$.\label{fig:zero-temperature internal energy}}
\end{figure}

\section{Equilibrium and thermal properties\label{sec:Equilibrium properties}}

Having introduced the simplified model, 
we present in this section the corresponding equilibrium properties
both at zero and finite temperature. These results show that our model captures the essential physics of $\mathrm{VO}_{2}$, justifying the later dynamics studies. We first determine
the ground-state phases and then study the phase transition from
the low-temperature distorted M1 phase to the high-temperature undistorted R phase.

\subsection{Ground-state phases}

We solve the model Hamiltonian \eqref{eq:Hamiltonian} using tensor
network methods within the Born-Oppenheimer approximation.
To determine the ground-state phases, we calculate the zero-temperature
adiabatic potential $\Phi_{\mathrm{eff}}(\mathbf{X})$ for each fixed
displacement $\mathbf{X}$, which is renormalized by the electronic
energy
\begin{equation}
\Phi_{\mathrm{eff}}(\mathbf{X})=\Phi(\mathbf{X})+\langle H_{\text{e-{\bf X}}}\rangle+\langle H_{\mathrm{e}}\rangle.
\end{equation}
Here the electronic energy (i.e., the last two terms)  is obtained by employing
the infinite density matrix renormalization group (iDMRG) method. The quarter filling is ensured in
the numerical simulation by introducing good quantum numbers. Due
to the $Z_{2}\times Z_{2}$ symmetry of the system under transformations
$X_{1,2}\to-X_{1,2}$ (domain inversion), we focus on the region with $X_{1},X_{2}>0$.

The results are shown in Fig.~\ref{fig:zero-temperature internal energy}.
There are two minima in the zero-temperature energy landscape (due to the $Z_{2}\times Z_{2}$ symmetry, the local minima at finite $\mathbf{X}$ are actually four-fold degenerate). One
local minimum is located at the origin point $X_{1}=X_{2}=0$ and corresponds
to the undistorted R phase. On the other hand, the global minimum is
located at $X_{1}\approx2.05$ and $X_{2}\approx1.65$, describing
the distorted M1 insulating ground state at zero temperature. From this, we conclude that the simplified
quasi-one-dimensional model \eqref{eq:Hamiltonian} captures the essential 
physics of $\mathrm{VO}_{2}$ and provides a good playground to qualitatively study its properties.

\subsection{Phase transition at finite temperature\label{subsec:phase transition at finite temperature}}

\begin{figure}
\includegraphics{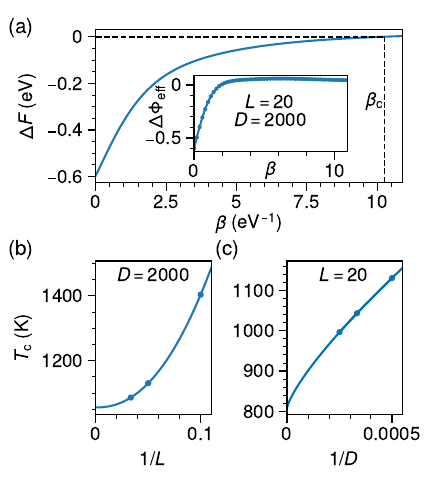}

\caption{Phase transition at finite temperature. (a) Free energy difference $\Delta F$ between the R and M1 phases as a function of the inverse temperature $\beta$ for system size $L=20$. The corresponding transition temperature $T_{\mathrm{c}}=1/k_{\mathrm{B}}\beta_{\mathrm{c}}\approx 1131\,\mathrm{K}$ is obtained by solving the equation $\Delta F(\beta)=0$. The insert shows the bare adiabatic potential difference $\Delta\Phi_{\mathrm{eff}}$, where the dots are obtained from the numerical simulation, while the line is the interpolation via polynomial function. We set the maximal bond dimension as $D=2000$ in this plot. (b) Finite size extrapolation for transition temperature $T_{\mathrm{c}}$ using the function $T_{\mathrm{c}}(L)=a+b/L^{c}$ with $a\approx 1055\,\mathrm{K}$, $b\approx 57292\,\mathrm{K}$, and $c\approx 2.2170$. The maximal bond dimension is fixed as $D=2000$. (c) Extrapolation of transition temperature $T_{\mathrm{c}}$ in the maximal bond dimension $D$ for system size $L=20$ using the function $T_{\mathrm{c}}(D)=a+b/D^{c}$ with $a\approx 810\,\mathrm{K}$, $b\approx 122306\,\mathrm{K}$, and $c\approx 0.7819$. Here the imaginary time step in the numerical simulation is set as $\delta\beta = 0.05\,\mathrm{eV}^{-1}$.
\label{fig:phase transition at finite temperature}}
\end{figure}

The key defining feature of $\mathrm{VO}_{2}$ is of course the transition
from the low-temperature M1 phase to the high-temperature R phase, but reproducing this thermal transition theoretically is nontrivial. 
Here we use the matrix product operator (MPO) time evolution technique~\citep{Zaletel2015,Paeckel2019}
in combination with the purification method~\citep{Verstraete2004}
to show that the simplified quasi-one-dimensional Hamiltonian \eqref{eq:Hamiltonian} can reproduce this finite-temperature 
phase transition and estimate the corresponding transition temperature.
We note that in this method the finite-temperature
state is obtained from the infinite-temperature state by imaginary
time evolution. To ensure the quarter filling, we use good quantum
numbers in the numerical simulation and start from a canonical infinite-temperature
ensemble with fixed particle number density and finite system size~\citep{Barthel2016}.
 While lattice entropy has been suggested to be important to the phase transition in $\mathrm{VO}_{2}$ previously~\citep{budai_metallization_2014,Wall2018}, for simplicity here we ignore this factor and focus on the electronic contribution. Including lattice entropy would, however, serve to further reduce the transition temperature.

We calculate the adiabatic
potential $\Phi_{\mathrm{eff}}$ at temperature $T$ and study the
temperature evolution of the free energies
\begin{equation}
F(\mathbf{X},T)=\Phi_{\mathrm{eff}}(\mathbf{X},T)-TS(\mathbf{X},T)
\end{equation}
for the two local minima we obtained at zero temperature~\citep{Grandi2020}. Since the imaginary
time evolution starts from the infinite temperature at which the entropies
$S_{\infty}$ are the same for both R and M1 phases, the entropy at temperature $T$ can be calculated through
\begin{equation}
S(\mathbf{X},T)=S_{\infty}-\int_{T}^{\infty}\mathrm{d}T'\,\frac{1}{T'}\frac{\partial\Phi_{\mathrm{eff}}(\mathbf{X},T')}{\partial T'}.
\end{equation}
Here the infinite-temperature entropy $S_{\infty}$ can be further eliminated
by considering the difference between the free energies of R
and M1 phases:
\begin{equation}
\Delta F(T)=\Delta\Phi_{\mathrm{eff}}(T)-T\Delta S(T)
\end{equation}
with $\Delta\Phi_{\mathrm{eff}}(T)\equiv\Phi_{\mathrm{eff}}(\mathbf{X}_{\mathrm{R}},T)-\Phi_{\mathrm{eff}}(\mathbf{X}_{\mathrm{M1}},T)$
and $\Delta S(T)=-\int_{T}^{\infty}\mathrm{d}T'\,(1/T')\partial\Delta\Phi_{\mathrm{eff}}(T')/\partial T'$. 
This quantity is what we are actually interested in.

The finite-temperature results are presented in Fig.~\ref{fig:phase transition at finite temperature},
where the imaginary time evolution is carried out with time step $\delta\beta=0.05\,\mathrm{eV}^{-1}$ ($\beta=1/k_{\mathrm{B}}T$ is the inverse temperature). Since the calculation for the entropy difference $\Delta S$ requires us to perform the differential and integration with respect to the temperature, we interpolate the adiabatic potential $\Delta \Phi_{\mathrm{eff}}(\beta)$ using polynomial functions. The obtained free energy difference from MPO time evolution with maximal bond dimension $D=2000$ for the system size $L=20$ is shown in Fig.~\ref{fig:phase transition at finite temperature}(a). The negative $\Delta F$ at high temperature indicates that the system is in the R phase, and there is a transition from the low temperature M1 phase ($\Delta F>0$), with the transition temperature $T_{\mathrm{c}}\approx 1131\,\mathrm{K}$ being identified by $\Delta F=0$. 

To obtain a more accurate estimation of $T_{\mathrm{c}}$, we perform the extrapolation in system size $L$ and maximal bond dimension $D$; see Figs.~\ref{fig:phase transition at finite temperature}(b) and \ref{fig:phase transition at finite temperature}(c). For the extrapolation in system size with fixed maximal bond dimension $D=2000$, the transition temperature for $L\to\infty$ is lowered by $\approx 76\,\mathrm{K}$ compared with the value for $L=20$. On the other hand, the maximal bond dimension $D$ has more notable influence on $T_{\mathrm{c}}$, as the changes of $\Delta F$ are very slow in low temperature, i.e., the temperature is more sensitive to $\Delta F$ in this region. For the system size $L=20$, the transition temperature for $D\to\infty$ is lowered by $\approx 321\,\mathrm{K}$ compared with the value for $D=2000$. Combining these two effects, we estimate the transition temperature for our used parameters as $T_{\mathrm{c}}\approx 734\,\mathrm{K}$.

We would like to mention that the exact transition temperature is hard to obtain using the MPO imaginary time evolution method. Although the estimated $T_{\mathrm{c}}$ is around a factor two higher when compared with the experimental value, our simplified model \eqref{eq:Hamiltonian} still qualitatively captures the phase transition from the low-temperature M1 phase to the high-temperature R phase. Therefore, we can use it to study the physics of light-induced phase transitions.  The difference may be due to the neglect of lattice entropy, which should be less important at very short times. Critically, as we show in Appendix \ref{sec:Quantum dynamics for our parameters and that with Tc being zero}, the light-induced dynamics for our parameters, which give $T_{\mathrm{c}}\approx 734\,\mathrm{K}$, are almost the same as compared to a system in which $T_{\mathrm{c}}=0$ . Hence, in this parameter range the deviation in transition temperature from the experimentally realized value will not affect the light-induced phase transitions qualitatively.

\section{Light-induced phase transitions\label{sec:Light-induced phase transitions}}

Having shown that our quasi-one-dimensional model can qualitatively 
capture the essential physics of $\mathrm{VO}_{2}$, especially the finite-temperature
phase transition, we next study in this section the light-induced phase transition from 
the initial M1 insulating phase to the long-time R metallic phase.

\subsection{Equations of motion\label{subsec:equations of motion}}

We excite the system using a pump pulse with the electric field
\begin{equation}
  \begin{aligned}
E_{\mathrm{pump}}(t) & = E_{0,\mathrm{pump}}e^{-(t-t_{0,\mathrm{pump}})^{2}/2\sigma_{\mathrm{pump}}^{2}} \\ &\quad\times\cos[\omega_{\mathrm{pump}}(t-t_{0,\mathrm{pump}})],
  \end{aligned}
\end{equation}
which is centered at time $t_{0,\mathrm{pump}}$ and has central frequency $\omega_{\mathrm{pump}}$
and temporal width $\sigma_{\mathrm{pump}}$. The pump pulse couples to the electronic degrees of freedom through the Peierls substitution $t_{\alpha}\to t_{\alpha}e^{\mathrm{i}A_{\mathrm{pump}}(t)}$ with the phase 
\begin{equation}
  \begin{aligned}
A_{\mathrm{pump}}(t)= & -(ed/\hbar)\int^{t}\mathrm{d}t'\,E_{\mathrm{pump}}(t')\\
= & A_{0,\mathrm{pump}}\sigma_{\mathrm{pump}}\exp(-\sigma_{\mathrm{pump}}^{2}\omega_{\mathrm{pump}}^{2}/2)\\
& \times[\mathrm{erf}(t_{-})-\mathrm{erf}(t_{+})],
\end{aligned}
\end{equation}
where $d$ is the lattice constant, $\mathrm{erf}(z)=(2/\sqrt{\pi})\int_{0}^{z}\mathrm{d}t\,e^{-t^{2}}$
is the error function, and we have $t_{\pm}=[\mathrm{i}\sigma_{\mathrm{pump}}^{2}\omega_{\mathrm{pump}}\pm(t-t_{0,\mathrm{pump}})]/\sqrt{2}\sigma_{\mathrm{pump}}$. With the pump pulse, the displacement $\mathbf{X}$ also becomes time-dependent due to the electron-lattice coupling.

We use the hybrid quantum-classical tensor-network method to simulate the dynamics of the system.
We start from the equilibrium M1 phase at zero temperature as numerous experiments have shown a negligible change in dynamics upon changing the initial temperature~\citep{Pashkin2011,DeLaPenaMunoz2023}. The time evolution of the
system can be decomposed into two parts, i.e., the quantum electronic and classical
lattice degrees of freedom. For the evolution of electronic state
$\vert\psi\rangle$, we can use the Born-Oppenheimer approximation
within each time step $\delta t$, i.e., the lattice distortions are
approximated as fixed, while the electronic degrees of freedom are
dynamic. Hence the electronic equation of motion is given by the Schr{\" o}dinger
equation and can be written as
\begin{equation}
\vert\psi(t+\delta t)\rangle=e^{-\mathrm{i}H[t,\mathbf{X}(t)]\delta t/\hbar}\vert\psi(t)\rangle,\label{eq:equation of motion for electrons}
\end{equation}
which can be simulated numerically by the infinite time-evolving block decimation (iTEBD) method. 
We note that we used the natural units in the numerical simulation, for which some of the simulation parameters like the time step $\delta t$ become irrational numbers in the international system of units.

On the other hand, for the lattice dynamics we use the classical approximation and invoke the Ehrenfest theorem for the lattice
degrees of freedom
\begin{equation}
M\frac{\mathrm{d}^{2}X_{i}}{\mathrm{d}t^{2}}=F_{i}(t)-\xi\frac{\mathrm{d}X_{i}}{\mathrm{d}t},\label{eq:equation of motion for lattice}
\end{equation}
where $M$ is the effective mass of ions, which is set as $25$ in
natural units, and $\xi$ is a damping coefficient used to model the lattice disordering observed in recent X-ray diffraction experiments~\citep{Wall2018,DeLaPenaMunoz2023,Johnson2023}. The forces $F_{i}$
are obtained through the Hellmann-Feynman theorem and explicitly read
\begin{equation}
\begin{aligned}F_{1}= & \frac{g}{2}\sum_{i=1,2}\cos(Qi)\langle\psi\vert n_{1,i}\vert\psi\rangle-\alpha X_{1}-2\beta_{1}X_{1}X_{2}^{2}\\
 & -\beta_{2}X_{1}(X_{1}^{2}-X_{2}^{2})-\gamma X_{1}(X_{1}^{2}+X_{2}^{2})^{2}
\end{aligned}
\label{eq:F1}
\end{equation}
and
\begin{equation}
\begin{aligned}F_{2}= & \frac{\delta}{2}X_{2}\sum_{i=1,2}\langle\psi\vert (n_{1,i}-n_{2,i})\vert\psi\rangle-\alpha X_{2}-2\beta_{1}X_{1}^{2}X_{2}\\
 & +\beta_{2}X_{2}(X_{1}^{2}-X_{2}^{2})-\gamma X_{2}(X_{1}^{2}+X_{2}^{2});
\end{aligned}
\label{eq:F2}
\end{equation}
see Appendix~\ref{sec:Hellmann-Feynman forces} for details.
With the equations of motion \eqref{eq:equation of motion for electrons}
and \eqref{eq:equation of motion for lattice}, both the light-induced structural and electronic dynamics
of $\mathrm{VO}_{2}$ can be simulated. % 

We note that treating the quantum electronic and classical lattice degrees of freedom separately within the Born-Oppenheimer approximation is similar to the real-time TD-DFT~\cite{Lian2018} used recently to model $\mathrm{VO}_{2}$~\citep{Xu2022,Liu2022}. However, in our hybrid quantum-classical tensor-network method, the electron-electron correlations are handled in a true many-body way, which enables capturing the full interaction effects.

\begin{figure}
\includegraphics{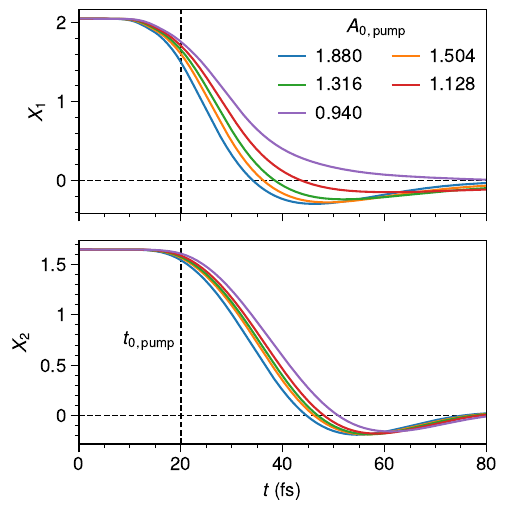}

\caption{Lattice dynamics for pump pulses with different amplitude $A_{0,\mathrm{pump}}$. 
Other parameters for the pulses are $\hbar\omega_{\mathrm{pump}}=1.5498\,\mathrm{eV}$, $\sigma_{\mathrm{pump}}=6\,\mathrm{fs}$,
and $t_{0,\mathrm{pump}}=20\,\mathrm{fs}$. Here we set the iTEBD time step as $\delta t=1.645\times 10^{-3}\,\mathrm{fs}$, and the maximal bond dimension is 1000. 
\label{fig:lattice dynamics for pulses with different strength and different damping coefficients}}
\end{figure}

\subsection{Photoinduced structural phase transition}
We first study the structural dynamics of $\mathrm{VO}_{2}$ induced by the pump pulse. 
We consider a pulse with wavelength $800\,\mathrm{nm}$, width $6\,\mathrm{fs}$, and centered
at $20\,\mathrm{fs}$. The electric field strength ranges from 0.5 to $1\,\mathrm{V/\mathring{A}}$. Note that for $\mathrm{VO}_{2}$ with lattice constant $d\approx 3\,\mathrm{\mathring{A}}$, 
the electric field of strength $E_{0,\mathrm{pump}} = 1\,\mathrm{V/\mathring{A}}$ corresponds to a Peierls
substitution phase of strength $A_{0,\mathrm{pump}}=1.88$. In the following, we will use $A_{0,\mathrm{pump}}$ to represent the pulse strength.

Fig.~\ref{fig:lattice dynamics for pulses with different strength and different damping coefficients} shows
the time evolution of lattice displacements $X_{1}$ and $X_{2}$ with damping coefficient $\xi=2$, chosen as a minimal value which removes the unphysical structural revivals beyond 100~fs delay, corresponding to the resolution of the best diffraction measurements~\citep{DeLaPenaMunoz2023}. 
The responses reflect well the ultrafast lattice dynamics observed in $\mathrm{VO}_{2}$. 
Especially, for the considered pulse strength the displacements $X_{1}$ and $X_{2}$ quickly transform to zero 
within the total simulation time $\sim 80\,\mathrm{fs}$, 
indicating the ultrafast photoinduced SPT from the distorted M1 phase to the undistorted R phase. 

However, there are also several interesting features in the structural dynamics at short time-scales not previously observed. The first is the overall timescale of the structural transition appears unrelated to the corresponding phonon modes when the system is excited below the transition threshold (see Appendix~\ref{sec:Lattice dynamics for systems excited below the transition threshold}). In particular, $X_{1}$ transforms in around the same time as the phonon mode would suggest (half period $\approx 21$~fs, crossing expected at $\approx 41$~fs), but $X_{2}$ transforms considerably faster (half period $\approx 44$~fs, crossing expected at $\approx 64$~fs). We note that the introduced damping slows the transition but is not relevant for the phonon, and so the transition likely outpaces the phonon even more than shown here. This suggests that, in contrast to assumptions in numerous studies~\citep{Cavalleri2001,Jager2017,Bionta2018}, the structural transition timescale is not limited by the normal phonon mode frequencies but in fact samples a significant portion of the nonlinear lattice potential. This nonlinearity means the common approach of using the timescale of the transition alone to assign a structural or electronic origin by comparison to known Raman modes could be highly misleading, not only for $\mathrm{VO}_{2}$ but for light-induced phase transitions generally. 

The second notable effect is that $X_{1}$ relaxes faster than $X_{2}$, i.e., the dimerization also relaxes prior to the tilt. This is broadly in-line with the two-step structural phase transition mechanism proposed by Baum \emph{et al.}~\citep{Baum2007}, but here occurs many orders of magnitude faster than the original proposal and is consistent with more recent diffraction measurements. This separation is also consistent with recent TD-DFT calculation~\citep{Xu2022}. 

Another remarkable feature of the lattice dynamics is that the displacements $X_{1}$ and $X_{2}$ undergo a transient revival with opposite sign for significant excitation levels. These findings can provide an explanation for the complex dynamics observed in Ref.~\citep{Brahms2024}, where the $a_{1g}$ band was found to exhibit a double-peak oscillatory structure at tens of femtoseconds in the time evolution. In that work, it was pointed out that the oscillation cannot be explained by the coherent electronic effect since the scattering time for electrons is much faster than this behavior, leaving these coherent lattice effects as the leading explanation. We note that for larger damping coefficient $\xi$, a stronger light pulse is required to observe the transient revival behavior of lattice displacements, but overall the dependence on pulse energy is quite weak, in contrast to recent TD-DFT calculation~\citep{Xu2022, Liu2022} and in agreement with X-ray diffraction measurements~\citep{Johnson2023}.

\subsection{Photoinduced electronic insulator-metal transition}

We now turn our attention to the IMT. Since it is hard to track the time-dependent occupations of single-particle states and the corresponding closure of the gap in a many-body method like iTEBD, here we instead study this phenomenon using the time-dependent optical conductivity and look for the collapse of the optical band gap. Given the knowledge of electronic wave function $\vert\psi(t)\rangle$
under the action of an external field $A(t)$, the temporal evolution
of the current, defined as $\langle J(t)\rangle=\langle\psi(t)\vert J(t)\vert\psi(t)\rangle$
with
\begin{equation}
J(t)\equiv\frac{\delta H(t)}{\delta A(t)}=-\mathrm{i}\sum_{a,\sigma,i}t_{a}[e^{\mathrm{i}A(t)}c_{a,\sigma,i}^{\dagger}c_{a,\sigma,i+1}-\mathrm{H.c.}],
\end{equation}
can be readily obtained, and we can extract the optical conductivity from this current.
For the systems at equilibrium, we set the external field
$A(t)$ to be a weak probe pulse $A_{\mathrm{probe}}(t)=A_{0,\mathrm{probe}}\exp[-(t-t_{0,\mathrm{probe}})^2/2\sigma^{2}_{\mathrm{probe}}]\cos[\omega_{\mathrm{probe}}(t-t_{0,\mathrm{probe}})]$,
and the corresponding current is denoted as $\langle J_{\mathrm{probe}}(t)\rangle$.
Since the wave function $\vert\psi(t)\rangle$ describes the influence
of $A_{\mathrm{probe}}$ on the ground state, the optical conductivity
at equilibrium can be calculated through
\begin{equation}
\sigma(\omega)=\frac{J_{\mathrm{probe}}(\omega)}{\mathrm{i}(\omega+\mathrm{i}\eta)LA_{\mathrm{probe}}(\omega)},\label{eq:optical conductivity}
\end{equation}
where $J_{\mathrm{probe}}(\omega)$ and $A_{\mathrm{probe}}(\omega)$
are the Fourier transformations of $\langle J_{\mathrm{probe}}(t)\rangle$
and $A_{\mathrm{probe}}(t)$, respectively. Numerically, a damping
factor $\exp(-\eta t)$ is introduced in the Fourier transformations,
which leads to the spectral broadening, but this factor can be set arbitrarily small commensurate to the step size of the simulation. 

We can extend this scheme to calculate the optical conductivity for
a nonequilibrium system driven by the pump pulse. To this end, we employ
the pump-probe based method proposed in Ref.~\citep{Shao2016}, where
the temporal evolution of the system is traced twice in order to identify
the response of the system with respect to the later probe pulse.
The procedure is as follows. First, the time-evolution process induced
by the pump pulse $A_{\mathrm{pump}}(t)$ in the absence of probe
pulse is evaluated, which describes the nonequilibrium development
of the system, and we have the current $\langle J_{\mathrm{pump}}(t)\rangle$.
Second, in addition to the pump pulse, we also introduce a weak 
probe pulse $A_{\mathrm{probe}}(t)$ centered at time $t_{*}$, which
leads to the current $\langle J_{\mathrm{total}}(t)\rangle$. The
subtraction of $\langle J_{\mathrm{pump}}(t)\rangle$ from $\langle J_{\mathrm{total}}(t)\rangle$
produces the variation of the current due to the presence of probe
pulse, i.e., $\langle J_{\mathrm{probe}}(t)\rangle$, with which the
time-dependent optical conductivity at time $t_{*}$ can be calculated
through Eq.~\eqref{eq:optical conductivity}.

\begin{figure}
\includegraphics{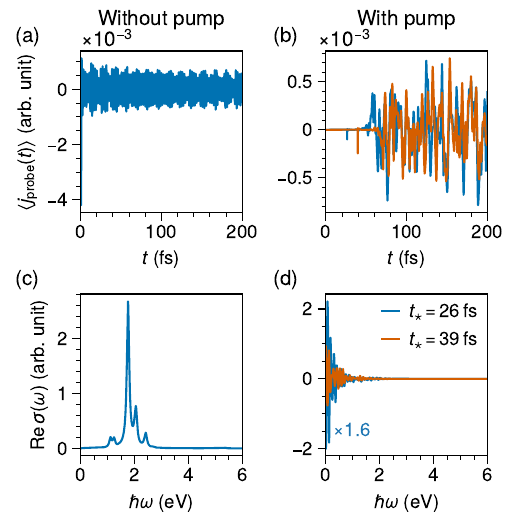}

\caption{Optical conductivity with and without the pump. (a), (b) The current
density $\langle j_{\mathrm{probe}}(t)\rangle$ due to the presence
of probe pulse with frequency $\hbar\omega_{\mathrm{probe}}=10\,\mathrm{eV}$, width
$\sigma_{\mathrm{probe}}=0.658\,\mathrm{fs}$, and amplitude $A_{0,\mathrm{probe}}=0.01$.
The center time $t_{0,\mathrm{probe}}$ of the probe pulse is $0.0658\,\mathrm{fs}$ for (a) and $26\,\mathrm{fs}$ (blue) or $39\,\mathrm{fs}$ (orange) for (b). 
(c), (d) The real part of the optical conductivity
obtained from $\langle j_{\mathrm{probe}}(t)\rangle$ shown in (a) and (b),
respectively. Here the time-dependent optical conductivity at $t_{*}=26\,\mathrm{fs}$ is rescaled for better visualization. 
For the pump pulse, the amplitude is chosen as $A_{0,\mathrm{pump}}=1.88$, and other parameters are the same as in 
Fig.~\ref{fig:lattice dynamics for pulses with different strength and different damping coefficients}. 
The time step for the iTEBD simulation is $\delta t=3.29\times 10^{-3}\,\mathrm{fs}$, and we set $\eta=0.075\,\mathrm{fs}^{-1}$ in the numerical Fourier transformation.
\label{fig:optical conductivity before and after the pump}}
\end{figure}

In Fig.~\ref{fig:optical conductivity before and after the pump},
we show the optical conductivity with and without the pump. For the
initial M1 phase at equilibrium, we apply a weak and narrow probe pulse of frequency $\hbar\omega_{\mathrm{probe}}=10\,\mathrm{eV}$
centered at $t_{0,\mathrm{probe}}=0.658\,\mathrm{fs}$ with width $\sigma_{\mathrm{probe}}=0.0658\,\mathrm{fs}$
and amplitude $A_{0,\mathrm{probe}}=0.01$ (i.e., a near-delta function), % (i.e., $t_{0,\mathrm{probe}}=1$ and $\sigma_{\mathrm{probe}}=0.1$ in the natural units),
which does not change the 
properties of the system qualitatively. Due to the finite time
step in the iTEBD numerical simulation, there is a small deviation from
zero for the current even in the absence of external fields. For this,
we also subtract this fictitious current from $\langle J_{\mathrm{probe}}(t)\rangle$.
The resulted current density induced by the probe pulse is shown in
Fig.~\ref{fig:optical conductivity before and after the pump}(a),
giving an optical conductivity with no amplitude at low frequencies and a first peak located at $\hbar\omega\approx1.1\,\mathrm{eV}$,
which identifies the insulating nature of the initial M1 phase; see
Fig.~\ref{fig:optical conductivity before and after the pump}(c).

On the other hand, the optical conductivity exhibits a sharply different
behavior following excitation by the pump pulse. 
Fig.~\ref{fig:optical conductivity before and after the pump}(b)
shows the current density $\langle j_{\mathrm{probe}}(t)\rangle$
for the probe pulses centered at $t_{0,\mathrm{probe}}=26\,\mathrm{fs}$ and $39\,\mathrm{fs}$,
times at which the lattice of $\mathrm{VO}_{2}$ is still distinct from the R structure
(cf. Fig.~\ref{fig:lattice dynamics for pulses with different strength and different damping coefficients}). 
Other parameters of the probe pulses remain the same as in the pump-free case.
The appearance of the Drude peak close to $\hbar\omega=0$ in the corresponding time-dependent 
optical conductivity {[}Fig.~\ref{fig:optical conductivity before and after the pump}(d){]}
shows the metallicity of the system by at least $t=26\,\mathrm{fs}$. This photoinduced electronic 
IMT is much faster than the SPT and can be considered 
as a quasi-instantaneous transformation, which is consistent with recent results~\citep{Baum2007,Kuebler2007,Pashkin2011,Wegkamp2014,Jager2017,Bionta2018}.
The decoupling nature of SPT and IMT 
in the light-induced nonequilibrium states also highlights the important role of electron-electron correlations
in driving the electronic transitions, which are indeed Mott-like instead of driven by the Peierls instability.
This effect is equally treated in the simplified quasi-one-dimensional model \eqref{eq:Hamiltonian} with the electron-lattice coupling 
and handled in a many-body way.

\section{Conclusion\label{sec:Conclusion}}

In conclusion, we have performed a tensor network study of the light-induced phase transitions in $\mathrm{VO}_{2}$. A simplified quasi-one-dimensional model was proposed to capture the corresponding essential physics, with all the important ingredients such as multi-orbital character, electron-lattice coupling, and electron-electron correlations being included. We shown that this model can qualitatively describe the equilibrium properties of $\mathrm{VO}_{2}$, such as the zero-temperature ground state phase diagram and finite-temperature phase transitions, which can provide insights into the studies of vanadium dioxide. 

Under the action of an ultrafast light pulse, we found a number of interesting structural and electronic behaviours. In agreement with a range of recent studies, we found that the electronic transition precedes the structural transitions~\citep{Brahms2024,Jager2017,Bionta2018}, supporting a Mott-like origin for the transition. However, we also found that the structure transforms faster than the harmonic phonon modes of the M1 phase, suggesting lattice nonlinearity is key in the SPT and the simple timescale arguments used to assign a structural or electronic nature to the transition from previous studies~\citep{Jager2017,Bionta2018,Cavalleri2004} do not necessarily apply for the more extreme case of light-induced phase transitions. This may have ramifications for light-induced phase transitions far beyond $\mathrm{VO}_{2}$.  Additionally, we found separate timescales for the evolution of dimerization and tilt distortions in the lattice dynamics, in broad agreement with older models of $\mathrm{VO}_{2}$~\citep{Baum2007} but here several orders of magnitude faster, in agreement with the timescales observed in more recent X-ray diffraction studies~\citep{DeLaPenaMunoz2023}.
Finally, we also observed a loss and subsequent restoration behavior of the structural displacements, which can provide an explanation for the complex dynamics recently found in the highest time-resolution studies to date~\citep{Brahms2024}. Future work will include systematic studies to find whether or not the IMT can be induced without also introducing the associated SPT~\citep{Morrison2014}, which would be a clear marker of the Mott behaviour, and examining to what degree the phase transition can be controlled using optical pulses~\citep{Johnson2023}. Our work sheds important light on the nature of the light-induced phase transition in $\mathrm{VO}_{2}$ at the shortest timescales, and challenges assumptions about signatures of decoupled electronic and structural phase transitions more generally. 

\begin{acknowledgments}
The iDMRG and finite temperature calculation were performed using the TeNPy library~\citep{Hauschild2018}, and the iTEBD was implemented based on the ITensor library~\citep{Fishman2022}.
A.~S.~J. acknowledges the support of the Ram{\' o}n y Cajal Program (Grant RYC2021-032392-I) and the Spanish AIE (projects PID2022-137817NA-I00 and EUR2022-134052), while IMDEA Nanociencia acknowledges support from the ``Severo Ochoa'' Programme for Centers of Excellence in R\&D (MICIN, CEX2020-001039-S).
T.~G. acknowledges funding by Gipuzkoa Provincial Council (QUAN-000021-01), by the Department of Education of the Basque Government through the IKUR strategy and through the project PIBA\_2023\_1\_0021 (TENINT), by the Agencia Estatal de Investigaci{\' o}n (AEI) through Proyectos de Generaci{\' o}n de Conocimiento PID2022-142308NA-I00 (EXQUSMI), by the BBVA Foundation (Beca Leonardo a Investigadores en F{\' i}sica 2023). The BBVA Foundation is not responsible for the opinions, comments and contents included in the project and/or the results derived therefrom, which are the total and absolute responsibility of the authors.  R.~W.~C. acknowledges support from the Polish National Science Centre (NCN) under the Maestro Grant No. DEC-2019/34/A/ST2/00081. 
ICFO group acknowledges support from: ERC AdG NOQIA; MCIN/AEI (PGC2018-0910.13039/501100011033, CEX2019-000910-S/10.13039/501100011033, Plan National FIDEUA PID2019-106901GB-I00, Plan National STAMEENA PID2022-139099NB-I00 project funded by MCIN/AEI/10.13039/501100011033 and by the ``European Union NextGenerationEU/PRTR'' (PRTR-C17.I1), FPI); QUANTERA MAQS PCI2019-111828-2); QUANTERA DYNAMITE PCI2022-132919 (QuantERA II Programme co-funded by European Union's Horizon 2020 program under Grant Agreement No 101017733), Ministry of Economic Affairs and Digital Transformation of the Spanish Government through the QUANTUM ENIA project call -- Quantum Spain project, and by the European Union through the Recovery, Transformation, and Resilience Plan -- NextGenerationEU within the framework of the Digital Spain 2026 Agenda; Fundaci{\' o} Cellex; Fundaci{\' o} Mir-Puig; Generalitat de Catalunya (European Social Fund FEDER and CERCA program, AGAUR Grant No. 2021 SGR 01452, QuantumCAT \textbackslash{} U16-011424, co-funded by ERDF Operational Program of Catalonia 2014-2020); Barcelona Supercomputing Center MareNostrum (FI-2023-1-0013); EU Quantum Flagship (PASQuanS2.1, 101113690); EU Horizon 2020 FET-OPEN OPTOlogic (Grant No 899794); EU Horizon Europe Program (Grant Agreement 101080086 -- NeQST), ICFO Internal ``QuantumGaudi'' project; European Union's Horizon 2020 program under the Marie-Sklodowska-Curie grant agreement No 847648; ``La Caixa'' Junior Leaders fellowships, ``La Caixa'' Foundation (ID 100010434): CF/BQ/PR23/11980043. Views and opinions expressed are, however, those of the author(s) only and do not necessarily reflect those of the European Union, European Commission, European Climate, Infrastructure and Environment Executive Agency (CINEA), or any other granting authority. Neither the European Union nor any granting authority can be held responsible for them.
U.~B. is also grateful for the financial support of the IBM Quantum Researcher Program.
\end{acknowledgments}

\appendix

\section{Hellmann-Feynman forces\label{sec:Hellmann-Feynman forces}}

In this Appendix, we derive the Hellmann-Feynman force $F_{i}$ for
the motion of lattice degrees of freedom. We denote the position of
the $i$-th ion as $(x_{1,i},x_{2,i})$ and the coordinate without distortion
as $(x^{(0)}_{1,i},x^{(0)}_{2,i})$, which are related to the displacement 
through $(X_{1},X_{2})=(x_{1,i}-x^{(0)}_{1,i}, x_{2,i}-x^{(0)}_{2,i})$.
With this, the lattice potential energy \eqref{eq:lattice potential energy}
and the lattice-electron coupling \eqref{eq:electron-lattice coupling} can be rewritten as
\begin{equation}
\begin{aligned}\Phi=\sum_{i} & \frac{\alpha}{2}\{[x_{1,i}-x^{(0)}_{1,i}]^{2}+[x_{2,i}-x^{(0)}_{2,i}]^{2}\}\\
 & +\frac{\beta_{1}}{4}\{2[x_{1,i}-x^{(0)}_{1,i}][x_{2,i}-x^{(0)}_{2,i}]\}^{2}\\
 & +\frac{\beta_{2}}{4}\{[x_{1,i}-x^{(0)}_{1,i}]^{2}-[x_{2,i}-x^{(0)}_{2,i}]^{2}\}^{2}\\
 & +\frac{\gamma}{6}\{[x_{1,i}-x^{(0)}_{1,i}]^{2}+[x_{2,i}-x^{(0)}_{2,i}]^{2}\}^{3}
\end{aligned}
\end{equation}
and
\begin{equation}
\begin{aligned}H_{\text{e-}\mathbf{X}}= & -g\sum_{i}[x_{1,i}-x^{(0)}_{1,i}]n_{1,i}\\
 & -\frac{\delta}{2}\sum_{i}[x_{2,i}-x^{(0)}_{2,i}]^{2}(n_{1,i}-n_{2,i}).
\end{aligned}
\end{equation}

\begin{figure}
\includegraphics{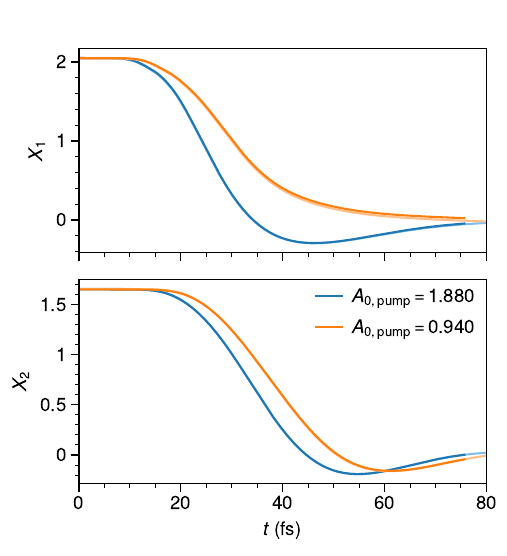}

\caption{Light-induced lattice dynamics for $g=0.52800\,\mathrm{eV}$ (dark colors) and $g=0.52715\,\mathrm{eV}$ (light colors) with other parameters of model \eqref{eq:Hamiltonian} being fixed. The simulations at $g=0.52715\,\mathrm{eV}$ are plotted to slightly longer final times to highlight the overlap. Here we set $\delta t=1.645\times 10^{-3}\,\mathrm{fs}$ and the maximal bond dimension is 1000 for the iTEBD simulation. Other parameters for the
pump pulses are the same as in Fig.~\ref{fig:lattice dynamics for pulses with different strength and different damping coefficients}. \label{fig:quantum dynamics for different parameters}}
\end{figure}

Therefore, the force acting on the $i$-th ion reads
\begin{equation}
\begin{aligned}f_{1,i}\equiv & -\biggl\langle\psi\biggl|\frac{\partial H}{\partial x_{1,i}}\biggr|\psi\biggr\rangle\\
= & g\langle\psi\vert n_{1,i}\vert\psi\rangle-\alpha[x_{1,i}-x^{(0)}_{1,i}]-2\beta_{1}[x_{1,i}-x^{(0)}_{1,i}]X_{2}^{2}\\
 & -\beta_{2}[x_{1,i}-x^{(0)}_{1,i}](X_{1}^{2}-X_{2}^{2})\\
 & -\gamma[x_{1,i}-x^{(0)}_{1,i}](X_{1}^{2}+X_{2}^{2})^{2}
\end{aligned}
\end{equation}
and
\begin{equation}
\begin{aligned}f_{2,i}\equiv & -\biggl\langle\psi\biggl|\frac{\partial H}{\partial x_{2,i}}\biggr|\psi\biggr\rangle\\
= & \delta[x_{2,i}-x^{(0)}_{2,i}]\langle\psi\vert (n_{1,i}-n_{2,i})\vert\psi\rangle-\alpha[x_{2,i}-x^{(0)}_{2,i}]\\
 & -2\beta_{1}[x_{2,i}-x^{(0)}_{2,i}]X_{1}^{2}+\beta_{2}[x_{2,i}-x^{(0)}_{2,i}](X_{1}^{2}-X_{2}^{2})\\
 & -\gamma[x_{2,i}-x^{(0)}_{2,i}](X_{1}^{2}+X_{2}^{2})^{2}.
\end{aligned}
\end{equation}

We consider a unit cell with two sites, for which we have
\begin{equation}
\begin{aligned}X_{1} & =\frac{x^{(0)}_{1,1}-x_{1,1}+x_{1,2}-x^{(0)}_{1,2}}{2}\\
 & =x^{(0)}_{1,1}-x_{1,1}=x_{1,2}-x^{(0)}_{1,2}
\end{aligned}
\end{equation}
and
\begin{equation}
\begin{aligned}X_{2} & =\frac{x^{(0)}_{2,1}-x_{2,1}+x_{2,2}-x^{(0)}_{2,2}}{2}\\
 & =x^{(0)}_{2,1}-x_{2,1}=x_{2,2}-x^{(0)}_{2,2}.
\end{aligned}
\end{equation}
Hence the corresponding Hellmann-Feynman force for the displacement $X_{i}$ is given by
\begin{equation}
F_{i}=\frac{f_{i,2}-f_{i,1}}{2},
\end{equation}
which leads to the expressions \eqref{eq:F1} and \eqref{eq:F2}.

\begin{figure}
\includegraphics{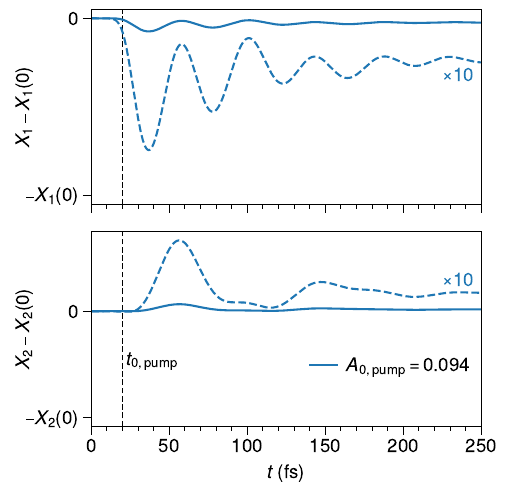}

\caption{Lattice dynamics for the weak pump pulse with $A_{0,\mathrm{pump}}=0.094$ below the transition threshold, for which the rescaled curves (dashed lines) are also added for better visualization. The oscillations, disregarding the initial displacement which is not directly proportional to the phonon oscillation, show a period of $\approx 43\,\mathrm{fs}$ ($88\,\mathrm{fs}$) for $X_{1}$ ($X_{2}$). Here the initial value of displacement is $X_{1}(0)\approx 2.05$ and $X_{2}(0)\approx 1.65$. We set $\delta t=1.645\times 10^{-3}\,\mathrm{fs}$ in the iTEBD simulation, and the maximal bond dimension is 1000. For the weak pump pulse below the transition threshold, the damping coefficient should be smaller than the cases with strong pulses and is set as $\xi=0.5$ in this plot. Other parameters for the pump pulses are the same as in Fig.~\ref{fig:lattice dynamics for pulses with different strength and different damping coefficients}. \label{fig:lattice dynamics for weak pulse}}
\end{figure}

\section{Comparison of quantum dynamics between our parameters and that with $T_{\mathrm{c}}=0$ \label{sec:Quantum dynamics for our parameters and that with Tc being zero}}

In this Appendix, we argue that although the transition temperature for our used parameters is a little high compared with the experimental value, the corresponding light-induced quantum dynamics does not change qualitatively. Since the exact transition temperature is hard to access for the MPO imaginary time evolution method, we achieve this by comparing the quantum dynamics of our parameters with that for parameters with $T_{\mathrm{c}}=0$. If the quantum dynamics for these two cases are close to each other, then it is similar for the parameters with $T_{\mathrm{c}}$ lying between zero and our estimated value.

Note that the parameters with transition temperature being zero correspond to the situation where the local minima in the zero-temperature energy landscape (cf. Fig.~\ref{fig:zero-temperature internal energy}) have the same value for the R and M1 phases. For our used parameters, the internal energy for the M1 phase is smaller than that for the R phase by $\approx 0.00137\,\mathrm{eV}$, which is quite small. Hence the parameters with $T_{\mathrm{c}}$ being zero will only change a little compared with our parameters. For this, we can fix other parameters and only change the electron-lattice coupling strength $g$. Then we have $g\approx 0.52715\,\mathrm{eV}$ for $T_{\mathrm{c}}=0$.

In Fig.~\ref{fig:quantum dynamics for different parameters}, we show the light-induced lattice dynamics of these two $g$'s starting from the M1 phase for different pulse amplitude $A_{0,\mathrm{pump}}$. Since the electron-phonon coupling strength $\delta$ is fixed, the time evolution of $X_{2}$ is indistinguishable visually for different $g$, and the main difference of lattice dynamics occurs in $X_{1}$. As we can see, the curves of $X_{1}$ are indeed close to each other for these two $g$'s, and the maximum difference is about $0.03$ when $A_{0,\mathrm{pump}}=0.940$, which is quite small compared with the initial value. Other pump pulses with stronger amplitude should have smaller difference. For the parameters with transition temperature between zero and our estimated $T_{\mathrm{c}}$, we would also expect that the difference of quantum dynamics will be further reduced. Hence in this region the light-induced phase transitions will not be affected qualitatively by the parameters and the corresponding transition temperature.

\section{Lattice dynamics for systems excited below the transition threshold \label{sec:Lattice dynamics for systems excited below the transition threshold}}
In this Appendix, we consider the case when the system is excited below the transition threshold. As an example, we show in Fig.~\ref{fig:lattice dynamics for weak pulse} the lattice dynamics for a weak pump pulse with $A_{0,\mathrm{pump}}=0.094$. % The displacements $X_{1}$ and $X_{2}$ only deviate from the initial position very slightly and stay finite at the long times. 
Compared with the strong pump pulse above the transition threshold (cf. Fig.~\ref{fig:lattice dynamics for pulses with different strength and different damping coefficients}), we find that the displacements $X_{1}$ and $X_{2}$ only deviate from the initial value very slightly, and the system undergoes a rapid change to a new equilibrium bond length around which it then oscillates with a period of 43~fs (88~fs) for $X_{1}$ ($X_{2}$).

\end{document}